\documentclass[aps,onecolumn,notitlepage,pra,superscriptaddress,nofootinbib,10pt]{revtex4-1}
\usepackage{graphicx}
\usepackage{graphics}
\usepackage{color}
\usepackage{amsmath,amssymb,amsthm}
\usepackage{csquotes}

\begin{document}

\title{Brief reply to ``Can gravity account for the emergence of classicality?''}
\author{Igor Pikovski}
\affiliation{ITAMP, Harvard-Smithsonian Center for Astrophysics, Cambridge, MA 02138, USA}
\affiliation{Department of Physics, Harvard University, Cambridge, MA 02138, USA}
\author{Magdalena Zych}
\affiliation{Centre for Engineered Quantum Systems, School of Mathematics and Physics, The University of Queensland, St Lucia, QLD 4072, Australia }
\author{Fabio Costa}
\affiliation{Centre for Engineered Quantum Systems, School of Mathematics and Physics, The University of Queensland, St Lucia, QLD 4072, Australia }
\author{\v{C}aslav Brukner}
\affiliation{Vienna Center for Quantum Science and Technology (VCQ), University of Vienna, Faculty of Physics, Boltzmanngasse 5, A-1090 Vienna, Austria}
\affiliation{Institute for Quantum Optics and Quantum Information (IQOQI), Austrian Academy of Sciences, Boltzmanngasse 3, A-1090 Vienna, Austria}


\begin{abstract}
In a series of comments, Bonder et al.\ criticized our work on decoherence due to time dilation~\cite{ref:Pikovski2015}. First the authors erroneously claimed that our results contradict the equivalence principle~\cite{bonderold}, only to ``resolve'' the alleged conflict in a second note~\cite{bondernew}. The resolution --  relativity of simultaneity -- was already explained in our reply~\cite{pikovski2015time}, which Bonder et al. now essentially reiterate. The newly raised points were also already extensively clarified in our note. The physical prediction of our work remains valid: systems with internal dynamics decohere if the superposed paths have different proper times.
\end{abstract}

\maketitle
\vspace{-2pt}

In a recent note, Bonder et al.\ \textquote{resolve an apparent conflict between [our] results and the equivalence principle} \cite{bondernew}. This apparent conflict was introduced by the same authors in their previous comment\footnote{The second arxiv version of this comment is quite different from the first one to which our reply \cite{pikovski2015time} referred. It also differs from the one submitted to Nature Physics. All versions state that our results contradict the equivalence principle.} claiming that our results \textquote{cannot be right in light of the \emph{equivalence principle} [because]
the situations of interest can be analyzed in a free falling frame [and] such scenarios cannot lead to decoherence, as, without gravity, there is nothing to cause it}~\cite{bonderold}. The claim is quite surprising considering that the effect we describe only depends on proper time and rest energy, which are coordinate-independent and quite compatible with the equivalence principle.
Indeed, the claim is incorrect. As Bonder et al.\ realize in their new note, what they interpreted as different reference frames
\textquote{are not describing the same physical situation from two perspectives, but two different situations}~\cite{bondernew}.
This precise point was extensively discussed in our answer to their concerns, sections IIIA-D of ref.~\cite{pikovski2015time}: \textquote{different observers will use different planes of simultaneity and thus will assign different states, simply because they are describing \emph{different physical situations}}\footnote{The similarity between their figure 4 in \cite{bondernew} and our figure 1 in \cite{pikovski2015time} is also quite evident, although these kind of figures and discussions are found in any exposition on basic relativity.}. What is important, however, is that once a physical situation is specified (e.g.\ an interference experiment), different reference frames give the same prediction, namely the one given in our work.

Bonder et al.\ further \textquote{emphasize that predictions regarding the observability of interference become relevant only in the context of concrete experimental settings}~\cite{bondernew}.
Indeed, we treat coherence as a physical property that is revealed in interference experiments, as emphasized throughout our works (for example in fig.~2 in \cite{ref:Pikovski2015}, see also ref.~\cite{ref:Zych2011}).
As interfering  paths have common initial and final points, no ambiguity related to the choice of equal-time surface arises.

Thus, despite the critical tone, in their most recent note Bonder et al.\ use our same explanations and reiterate the central prediction of our work: systems with internal dynamics decohere if the superposed paths have different proper times.
A few points of confusion remain by Bonder et al., which we already resolved in ref.~\cite{pikovski2015time}. First, in the use of the word ``universal'': As explained in our works, it means that time dilation affects all composite systems, not that the effect is inevitable. Second, in the role of decoherence: As commonly understood in the literature, it suppresses quantum effects in experiments, which thus appear classical for all practical purposes, it does not turn quantum states into ``proper mixtures''.
Third, in the mass superselection rule: it has no relevance for our work, as we are discussing relativistic effects.
Finally, Bonder et al.\ question in which sense gravitation is related to the effect: it is in providing the gravitational time dilation responsible for decoherence, as the title of our work suggests. These and further concepts underlying our work are discussed in ref.~\cite{pikovski2015time}.

\end{document}